\documentclass{article}
\usepackage{srcltx}
\usepackage[english]{babel}
\usepackage{amsmath}
\usepackage{amssymb}
\usepackage{amsfonts}
\usepackage{latexsym}
\usepackage{textcomp}
\usepackage{appendix}
\usepackage{multirow}
\usepackage{booktabs}
\usepackage{rotating}
\usepackage{afterpage}
\usepackage{pdflscape}
\usepackage{url}
\usepackage{array}
\usepackage{marvosym}
\usepackage{graphics}
\usepackage{graphicx}
\usepackage{natbib}
\usepackage[table]{xcolor}
\usepackage{epsfig}
\usepackage{float}
\usepackage{authblk}

\usepackage[caption=false]{subfig}

\usepackage[margin=1.5in]{geometry}
\makeatletter
\renewcommand*{\@fnsymbol}[1]{\ensuremath{\ifcase#1\or *\or \mathparagraph\or \dagger\or \ddagger\or
   \mathsection\or  \|\or **\or \dagger\dagger
   \or \ddagger\ddagger \else\@ctrerr\fi}}
\makeatother
\title{The link between Bitcoin and Google Trends attention}

\author[1]{Nektarios Aslanidis \thanks{nektarios.aslanidis@urv.cat}}
\author[2]{Aurelio F. Bariviera \thanks{aurelio.fernandez@urv.cat}}
\author[1,2]{\'Oscar G. L\'opez}
\affil[1]{\scriptsize  Universitat Rovira i Virgili, Department d'Economia, ECO-SOS, Reus, Spain}
\affil[2]{\scriptsize  Universitat Rovira i Virgili, Department of Business, Reus, Spain}

\begin{document}
\maketitle

\begin{abstract}
This paper shows that Bitcoin is not correlated to a general uncertainty index as measured by the Google Trends data of \cite{Castelnuovo2017}. Instead, Bitcoin is linked to a Google Trends attention measure specific for the cryptocurrency market. First, we find a bidirectional relationship between Google Trends attention and Bitcoin returns up to six days. Second, information flows from Bitcoin volatility to Google Trends attention seem to be larger than information flows in the other direction. These relations hold across different sub-periods and different compositions of the proposed Google Trends Cryptocurrency index.

\end{abstract}


\section{Introduction \label{sec:intro}}

Since its creation in 2009, Bitcoin has gained a growing attention among investors, researchers and policy makers. The first advocates were libertarians critical to the global financial crisis of 2008. These investors saw blockchain as a mechanism to bypass the traditional financial system, which was severely criticized as its lax regulation was deemed to lead to the crisis. A second wave of Bitcoin enthusiasts were speculators, who saw in Bitcoin (and in newly minted cryptocurrencies) high-yield investment opportunities. A third wave of market participants were financial institutions, which aimed to introduce blockchain technology in their industry and offer investors more secure platforms for cryptoinvestment. At the same time, governments begun to worry about the potential negative effects of cryptocurrencies. Several countries have been introducing regulations (e.g., tax laws, anti-money laundering/anti-terrorism financing laws. See \cite{GlobalLegalResearchCenter2018}) and issuing warnings about the high risk of this type of investment \citep{FT2021}. Finally, a fourth wave of cryptocurrency market players currently taking place, is related to the so-called Central Bank Digital Currencies \citep{FernandezVillaverde2020}. For a detailed review on the evolution and current state of research on cryptocurrencies, we refer to \cite{Corbet2019} and \cite{BarivieraMerediz2021}, among others. 

At the same time, the increasing digitalization of the economy has left a digital footprint that, in certain way, reveals preferences, tastes, or consumption habits. Given that cryptocurrencies are digitally native assets, investors tend to gather market information mainly through the internet (social networks, specialized forums, etc.). Specifically, Google searches tend to signal investors' attention. \cite{URQUHART2018} is one of the earliest papers to relate cryptocurrency's market attention with Google Trends, finding that realized volatility, volume and returns influence future search for the term 'Bitcoin'. Subsequently, \cite{SHEN2019118} point out that the number of tweets is a significant driver of Bitcoin trading volume and realized volatility.

Other researchers have used news-based uncertainty indices to assess the impact of uncertainty on Bitcoin. \cite{DEMIR2018145} finds that the Economic Policy Uncertainty (EPU) index is negatively associated with Bitcoin daily returns. \cite{Walther2019}, using a GARCH-MIDAS framework, finds that Global Real Economic Activity fares well in cryptocurrency volatility forecasting. In a similar vein, \cite{Fang2020} reports a significant impact of News Implied Volatility (NVIX) on long-term cryptocurrency volatility. Meanwhile, \cite{Aysan2019} detects significant predictive power of the Geopolitical Risk (GPR) index for both Bitcoin returns and volatility. More recently, \cite{Lucey2021} uses weekly data to construct cryptocurrency uncertainty indices based on a variety of news pieces from LexisNexis Business database. The authors carry out a historical decomposition and relate their cryptocurrency uncertainty indices to major economic and political events.

The aim of the present paper is to explore to what extent investors' attention to the cryptocurrency market is captured by a set of keywords as measured by Google Trends. Compared to Twitter (where access is limited in time) or to LexisNexis (a subscription-based service), Google Trends is freely available. In addition, Google Trends is simple to obtain and potentially reflects the attention of a broader profile of investors. 

Overall, our contribution to the literature is as follows. First, we construct a Google Trends cryptocurrency index to capture market attention. Second, we find that the Google Trends Uncertainty (GTU) index proposed by \cite{Castelnuovo2017} does not prove useful in reflecting cryptocurrency market attention. Third, we show that there are important information flows from Google Trends to the cryptocurrency market and viceversa, reflecting a recurring dialog between the market attention and investors' interests. Finally, cryptocurrency market attention is well captured by a handful of keywords  such as Bitcoin, BTC, blockchain, crypto, cryptocurrency.

The rest of the paper is organized as follows: Section \ref{sec:index-construction} details the proposed
Google Trends Cryptocurrency index; section \ref{sec:methods} briefly describes the key methodologies used in the paper; section \ref{sec:data-results} describes our data set and discusses the main finding. Finally, Section \ref{sec:conclusion} concludes.

\section{Construction of Google Trends Cryptocurrency index\label{sec:index-construction}}

Following \cite{Castelnuovo2017}, we update their Google Trends Uncertainty (GTU) index over the period 2015-2021. In addition, we construct a Google Trends Cryptocurrency (GTC) index, using a set of cryptocurrency-oriented keywords. It is reasonable to postulate that cryptocurrency investors gather information mainly through the internet. Even though there are several search engines (e.g., Google, Yahoo, Bing, Ask), Google clearly dominates the market. According to \cite{Statista} the worldwide search market share of Google is 86.6\%. Thus Google Trends could be used as a reliable measure for online searches. 

The keywords that constitute our GTC index are selected using a bibliometric analysis of scientific papers in line with \cite{MeredizBariviera2019}. The full set is composed by 38 keywords. In order to check the robustness of our results, we reduce the number of keywords, leaving only the ones that we consider more closely related to Bitcoin economics, while dropping more technical words such as 'hash', 'hard fork', 'proof of work', etc.  The list of the full set of keywords, as well as the three subsets of keywords of the index are detailed in the appendix. After obtaining the daily Google Trend index for each keyword, we compute the arithmetic mean for all the keywords, constituting the daily GTC index. 

\begin{figure}[H]
\centering
\subfloat[][GTC index]{\includegraphics[width = 0.45\textwidth]{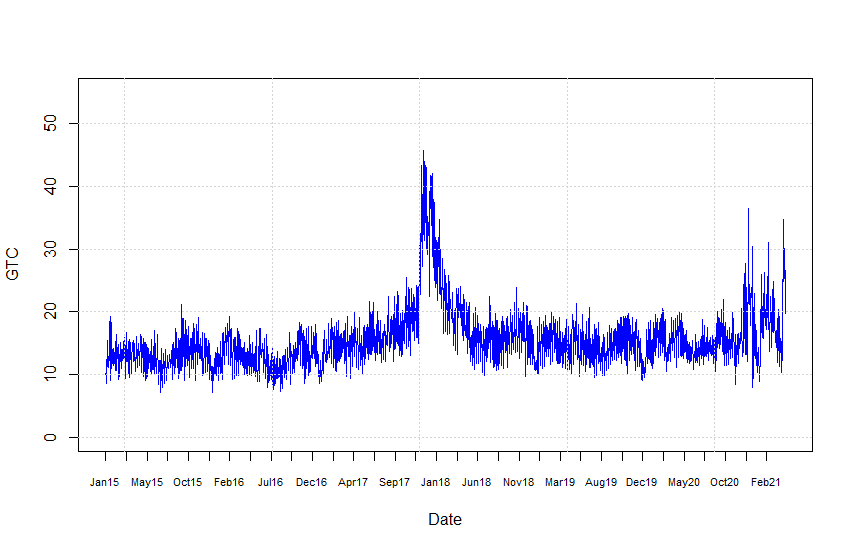}}
\subfloat[][GTU index]{\includegraphics[width = 0.45\textwidth]{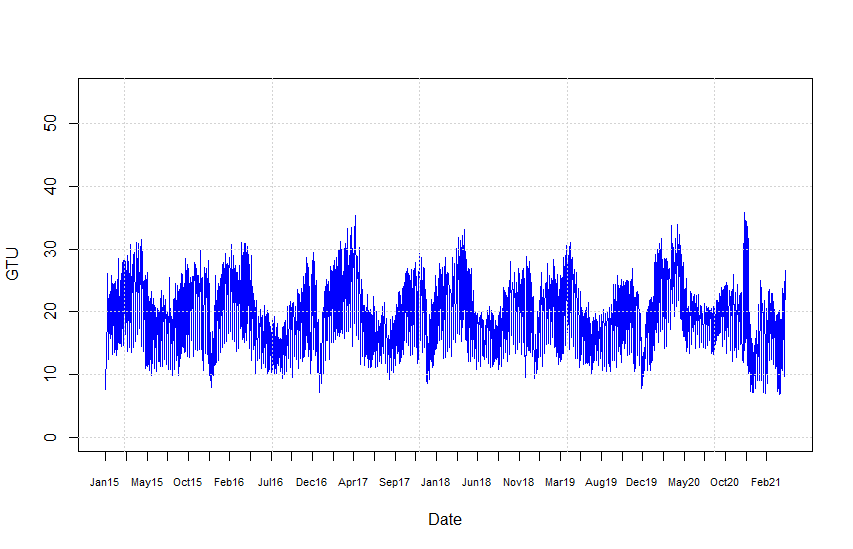}}
\caption{Google Trends attention indices: GTC and GTU \citep{Castelnuovo2017}.}
\label{fig:GTU-GTC}
\end{figure}

\section{Methods \label{sec:methods}}

We measure information linkages between Bitcoin and market attention by means of Shannon Transfer Entropy (for details, see \cite{Dimpfl2013,Dimpfl2018}). Shannon Transfer Entropy is a flexible, non-parametric method designed to overcome some of the limitations of Granger causality (the assumed linearity assumption). 

Transfer Entropy is a measure based on the Kullback-Liebler distance of transition probabilities, and allows not only to determine the direction of information flows, but more importantly to quantify the strength of those flows.   

Let consider two processes $I$ and $J$, with marginal probability distributions $p(i)$ and $p(j)$, and joint probability distribution $p(i,j)$, the Shannon transfer entropy (TE) can be defined as:
\begin{equation}
T_{J \rightarrow I}(k,l) = \sum_{i,j} p\left(i_{t+1}, i_t^{(k)}, j_t^{(l)}\right) \cdot log \left(\frac{p\left(i_{t+1}| i_t^{(k)}, j_t^{(l)}\right)}{p\left(i_{t+1}|i_t^{(k)}\right)}\right),
\end{equation}
where $T_{J \rightarrow I}$ is a measure of the information conveyed from $J$ to $I$. Considering that TE could be a biased estimator of the information transfer, \cite{Marschinski2002} proposed a modified metric, by removing the information produced by shuffled realization of the explanatory process as:
\begin{equation}
ETE_{J \rightarrow I}(k,l)=  T_{J \rightarrow I}(k,l)- T_{J_{\text{shuffled}} \rightarrow I}(k,l)
\end{equation}
The Effective Transfer Entropy (ETE) shows not only the direction, but also the quantity of information transmited from one process to the other. 

\section{Empirical Analysis \label{sec:data-results}}
\subsection{Data}
This paper uses daily data on Google Trends and Bitcoin prices. We calculate the Google Trends Uncertainty (GTU) index using the keywords of \cite{Castelnuovo2017}, as well as our Google Trends Cryptocurrency (GTC) index proposed in Section \ref{sec:index-construction}. Bitcoin daily data is used to compute the logarithmic return and \cite{Parkinson1980} volatility\footnote{We also computed \cite{GarmanKlass} volatility. Although not displayed in the paper due to space considerations, results are similar.}. For replication purposes, data used in this paper is available online along this paper. 

We focus mainly on Bitcoin, since cryptocurrency market linkages (both in returns and volatilities) have become very strong in recent years \citep{Aslanidis2021}.

The empirical results of this section are obtained using the GTC index with five keywords (Subset 2), but they are robust to a different selection of keywords. For details of the different selection of keywords, see the appendix.

\subsection{Results}

\begin{figure}[!htbp]
\centering
\subfloat[][Daily return]{\includegraphics[width = 0.45\textwidth]{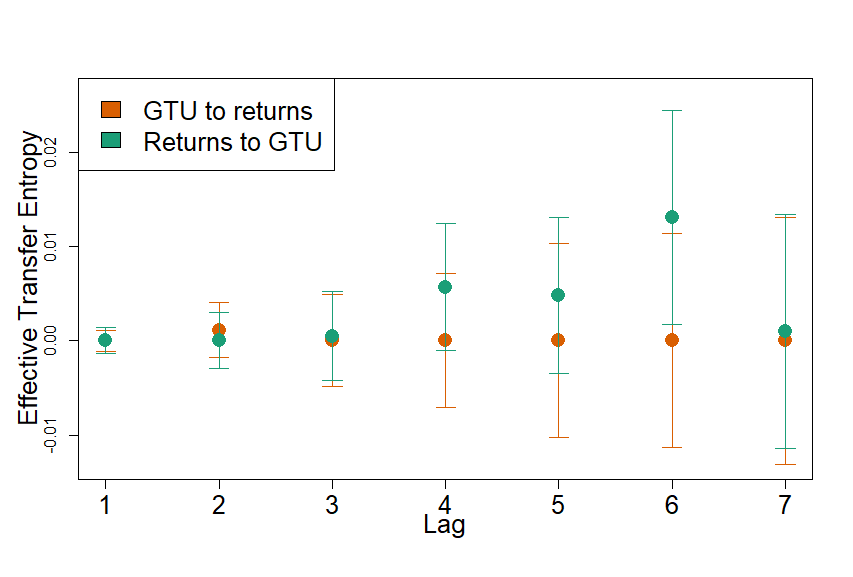}}
\subfloat[][Volatility (Parkinson)]{\includegraphics[width = 0.45\textwidth]{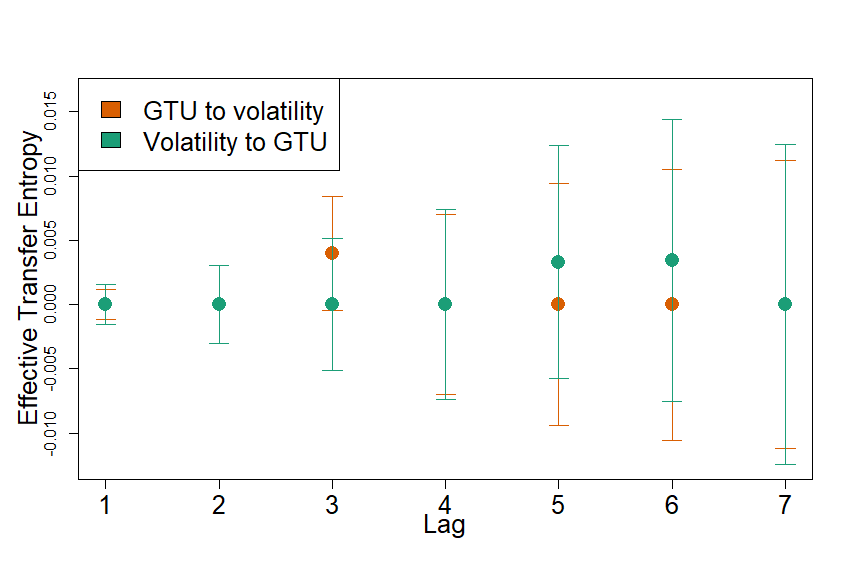}}
\caption{Effective Transfer Entropy (ETE) between Google Trends Uncertainty (GTU) and return (a) or volatility (b) using different lags.}
\label{fig:GTUlags}
\end{figure}

We explore by means of Transfer Entropy, the information exchange between the Bitcoin market and the Google Trends uncertainty (GTU) index. We use first differences of the Google Trends data to ensure stationarity. Table \ref{tab:ETEgtu} shows that information flows between Bitcoin and GTU are not statistically significant, which indicates a detachment of cryptocurrencies from the general macroeconomic environment. This result is in line with previous findings (\cite{CORBET201828} and \cite{Aslanidis2019}), who report that major cryptocurrencies are rather isolated from traditional assets such as gold, stocks or bonds. Notice our finding is robust to a selection of different lag lengths, as observed in Figure \ref{fig:GTUlags}.

\begin{table}[!htbp]
  \centering
  \caption{Transfer Entropy between Google Trends Uncertainty (GTU) and Bitcoin (return and volatility)}
    \begin{tabular}{lrrrr}
    \toprule
    Direction & \multicolumn{1}{l}{TE} & \multicolumn{1}{l}{ETE} & \multicolumn{1}{l}{Std.Err.} & \multicolumn{1}{l}{p-value} \\
    \midrule
    GTU\textrightarrow Return & 0.0036 & 0.0000 & 0.0013 & 0.2267 \\
    Return\textrightarrow GTU & 0.0036 & 0.0006 & 0.0014 & 0.5067 \\
          &       &       &       &  \\
    GTU\textrightarrow Volatility & 0.0022 & 0.0000 & 0.0013 & 0.6067 \\
    Volatility \textrightarrow GTU & 0.0031 & 0.0004 & 0.0015 & 0.4967 \\
    \bottomrule
    \end{tabular}%
  \label{tab:ETEgtu}%
\end{table}%

A different picture emerges, however, when analyzing Bitcoin returns/volatility with respect to the Google Trends Cryptocurrency (GTC) index. Table \ref{tab:ETEgtc} displays the results of Transfer Entropy using just the first lag. As seen, there is a reciprocal flow of information between Bitcoin and GTC. We observe that the amount of information emitted and received between Bitcoin returns and market attention is similar. However, there is more information leaked from Bitcoin volatility to GTC than in the other direction. 

\begin{figure}[!htbp]
\centering
\subfloat[][Daily return]{\includegraphics[width = 0.45\textwidth]{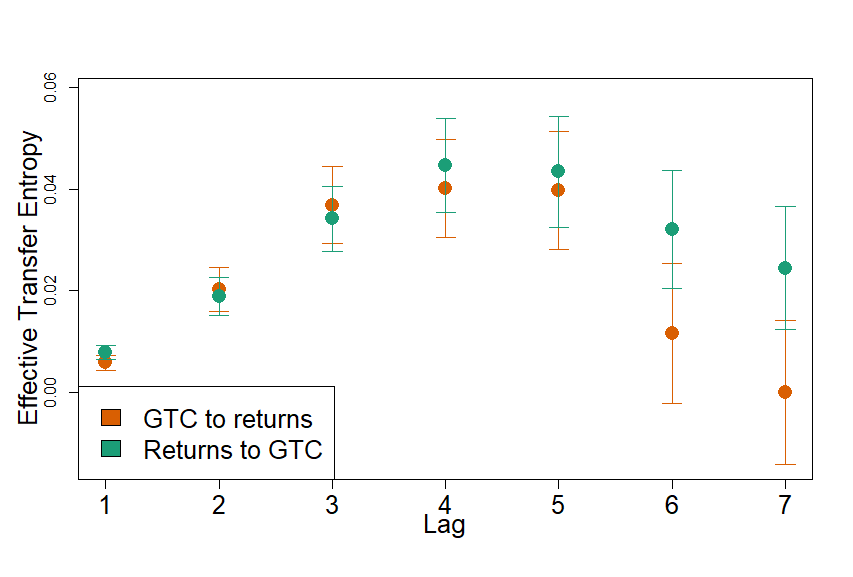}}
\subfloat[][Volatility (Parkinson)]{\includegraphics[width = 0.45\textwidth]{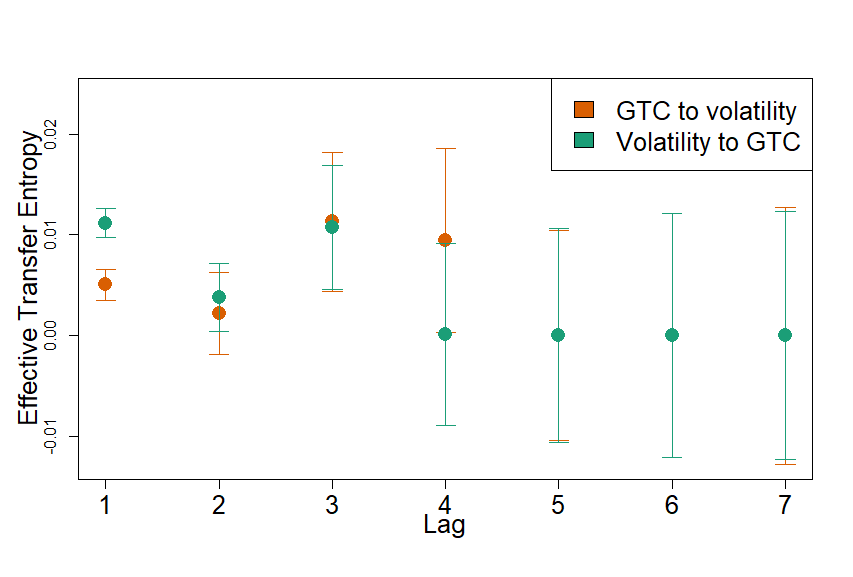}}
\caption{Effective Transfer Entropy (ETE) between Google Trends Cryptocurrency (GTC) index and return (a) or volatility (b) using different lags.}
\label{fig:GTClags}
\end{figure}

Another important result is that the interdependence between GTC and returns is significant for up to six days (see Figure \ref{fig:GTClags}). On the contrary, the transfer of information between Bitcoin volatility and GTC holds strong for up to three days. This implies that news are  absorbed by the market,
although large price swings produce stronger market attention during several days. 

\begin{figure}[!htbp]
\centering
\subfloat[][Daily return]{\includegraphics[width = 0.45\textwidth]{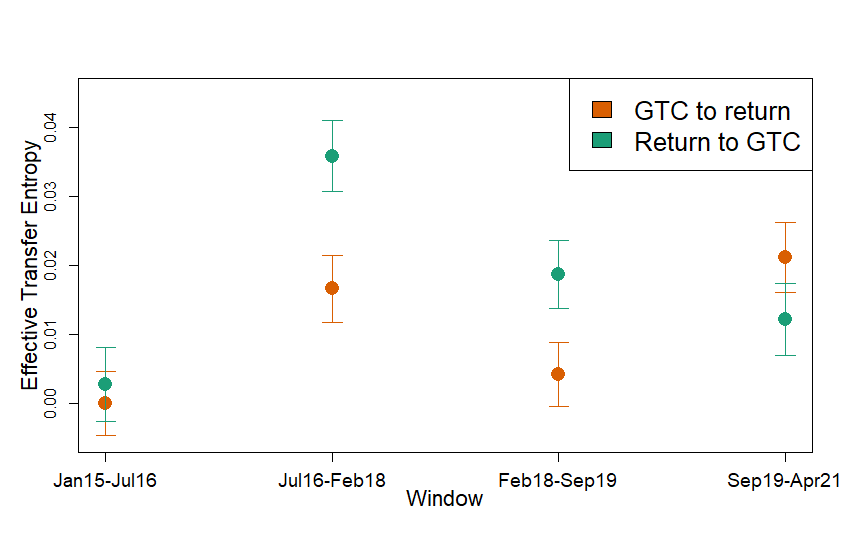}}
\subfloat[][Volatility (Parkinson)]{\includegraphics[width = 0.45\textwidth]{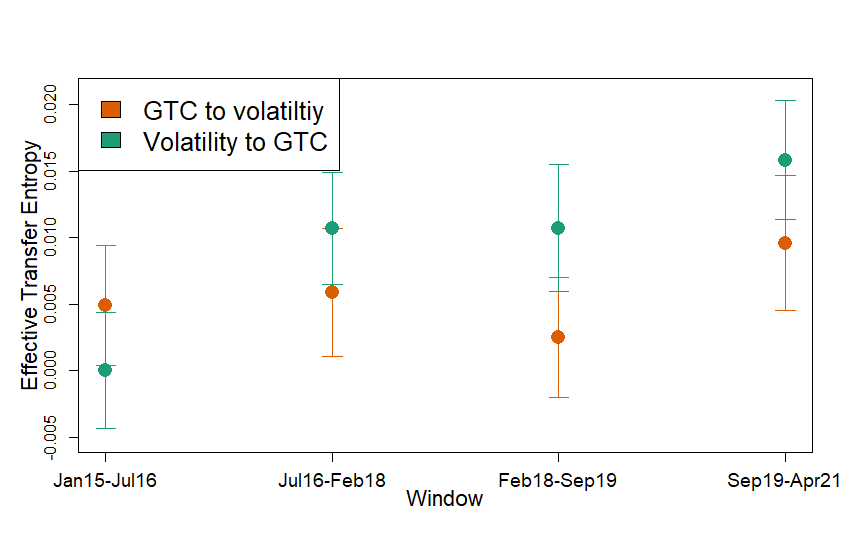}}
\caption{Effective Transfer Entropy (ETE) between Google Trends Cryptocurrency (GTC) index and return (a) or volatility (b) using non overlapping windows.}
\label{fig:GTCwindows}
\end{figure}

\begin{figure}[!htbp]
\centering
\includegraphics[width = 0.5\textwidth]{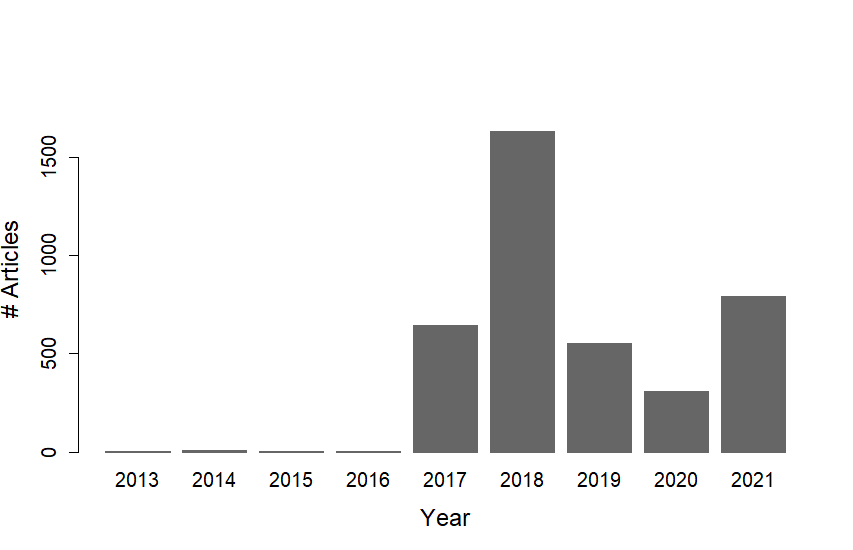}
\caption{Number of articles under the subject `cryptocurrency markets' published per year in Dow Jones Newswires, MarketWatch, The Wall Street Journal and Barron's. Source: Factiva.}
\label{fig:factiva}
\end{figure}

To analyze the information transfer over time, we divide our sample into four non-overlapping windows of 575 observations each. Results of ETE using the first lagged value of the variables are reported in Figure \ref{fig:GTCwindows}. We observe that in the first window (from January 2015 to July 2016) there is no information transfer between returns or volatilities and GTC. We should recall that until 2016 there was less public knowledge of cryptocurrencies. It was precisely in 2017 when Bitcoin price rose from \$ 800 to \$ 20000, and cryptocurrencies begun gaining attention in social networks and mainstream newspapers. Figure \ref{fig:factiva} clearly shows that since 2017 prestigious sources such as Dow Jones Newswires, MarketWatch, The Wall Street Journal, and Barron's have been publishing articles within the subject `Cryptocurrency Markets' \citep{Factiva}. News made cryptocurrencies popular, encouraging high-yield seeking investors to gather information about this new type of financial asset, generating this ``dialog'' between online searches and cryptocurrency market profitability and risk metrics.   

\begin{table}[htbp]
  \centering
  \caption{Transfer entropy between GTC and Bitcoin (return and volatility)}
    \begin{tabular}{lrrrr}
    \toprule
    Direction & TE    & ETE   & Std.Err. & p-value \\
    \midrule
    GTC \textrightarrow Return & 0.0098 & 0.0058 & 0.0015 & 0.0025 \\
    Return \textrightarrow GTC & 0.0121 & 0.0079 & 0.0013 & 0.0000 \\
          &       &       &       &  \\
    GTC \textrightarrow Volatility & 0.0083 & 0.0051 & 0.0015 & 0.0100 \\
    Volatility \textrightarrow GTC & 0.0155 & 0.0111 & 0.0014 & 0.0000 \\
    \bottomrule
    \end{tabular}%
  \label{tab:ETEgtc}%
\end{table}%

\section{Conclusion\label{sec:conclusion}}

We confirm that the cryprocurrency market is rather detached from the general macroeconomic environment as proxied by the Google Trends Uncertainty data \cite{Castelnuovo2017}. Instead, the proposed Google Trends Cryptocurrency index conveys important information flows to (and receives feedback from) the cryptocurrency market.

Information transfer between Google Trends and daily returns is found to be bidirectional and to last for up to six days. As for Bitcoin volatility news are more rapidly absorbed, as the significance of information flows vanishes after three days. Another worth noting result is that information flows from Bitcoin volatility to Google Trends data are larger than vice versa.

When analyzing data in temporal sub-samples, we detect that information flows are not significant at the start of our sample (January 2015-July 2016), but become so after 2016. This could be explained by the greater attention and consolidation of the cryptocurrency market as an alternative investment vehicle. 

Our results are robust to different composition of the Google Trends Cryptocurrency (GTC) index, to different lag lengths, and sub-periods. 

Recent moves by investment banks into cryptocurrencies (e.g., Morgan Stanley, Goldman Sachs) and the approval of Bitcoin ETFs in Canada show institutional interest in the market is gaining momentum. More recently, the \cite{BIS2021} started a consultative process to gather opinions on the possibility of commercial banks holding Bitcoin and other digital assets. Although our research does not detect any relationship between Bitcoin and the general macroeconomy, the institutionalization of cryptocurrencies could make their way into the traditional financial ecosystem. Overall, our research has important implications for fund management and policy making, as it provides important information for portfolio design and rebalancing.

\bibliographystyle{apalike}
\bibliography{biblio}

\appendix

\section{Sets of words to construct the Google Trends Cryptocurrency Attention (GTC) Index}

\begin{table}[H]
  \centering
  \caption{Sets of words considered in the Google Trends Cryptocurrency (GTC) Attention Index}
    \begin{tabular}{clllr}
    \toprule
    \multirow{10}[2]{*}{Full set} & AES256 & crypto crash & miner & \multicolumn{1}{l}{Mount Gox} \\
          & altcoin & cryptocurrency & minted & \multicolumn{1}{l}{Mt Gox} \\
          & anonymity & cryptography & public key & \multicolumn{1}{l}{Mt. Gox} \\
          & Bitcoin & digital assets & ripple & \multicolumn{1}{l}{private key} \\
          & block producer & distributed ledger & satoshi & \multicolumn{1}{l}{Proof of Authority} \\
          & blockchain & ethereum & soft fork & \multicolumn{1}{l}{Proof of Burn} \\
          & BTC   & hard fork & stablecoin & \multicolumn{1}{l}{Proof of Stake} \\
          & coin  & hash  & tether & \multicolumn{1}{l}{Proof of Work} \\
          & consensus & hashing & token &  \\
          & crypto & ICO   & virtual currency & \multicolumn{1}{l}{tether} \\
    \midrule
    \multirow{6}[2]{*}{Subset 1} & anonymity & crypto & miner & \multicolumn{1}{l}{token} \\
          & Bitcoin & cryptocurrency & minted & \multicolumn{1}{l}{private key} \\
          & blockchain & ICO   & public key & \multicolumn{1}{l}{Proof of Work} \\
          & BTC   & cryptography & ripple &  \\
          & coin  & ethereum & satoshi &  \\
          & consensus & hash  & stablecoin &  \\
    \midrule
 \multirow{3}[2]{*}{Subset 2} & Bitcoin & crypto & ethereum & \multicolumn{1}{l}{tether} \\
          & blockchain & cryptocurrency & ripple &  \\
          & BTC   & cryptography & satoshi &  \\
        \midrule
      
          \multirow{2}[2]{*}{Subset 3} & Bitcoin & BTC   & cryptocurrency &  \\
          & blockchain & crypto &       &  \\
   
    \bottomrule
    \end{tabular}%
  \label{tab:word-set}%
\end{table}%

\end{document}